\documentclass[twocolumn]{article}

\usepackage{amsmath}
\usepackage{tabularx}
\usepackage{comment}
\usepackage[ruled, vlined, linesnumbered]{algorithm2e}
\usepackage{algpseudocode}
\usepackage{latexsym}
\usepackage{amsfonts}
\usepackage{listings}
\usepackage{tikz} 
\usepackage{mathrsfs} 
\usepackage{subfig}
\usepackage{url}

\AtBeginDocument{%
  \providecommand\BibTeX{{%
    \normalfont B\kern-0.5em{\scshape i\kern-0.25em b}\kern-0.8em\TeX}}}

\graphicspath{./image/}

\newtheorem{defi}{Definition}
\newtheorem{problem}{Problem}

\makeatletter
\newcommand{\figcaption}[1]{\def\@captype{figure}\caption{#1}}
\newcommand{\tblcaption}[1]{\def\@captype{table}\caption{#1}}

\newcommand{\queryattr}{{\tau_Q}}
\newcommand{\workflowattr}{{\tau_N}}
\newcommand{\workflowgraph}{{W_N}} 
\newcommand{\x}{x} 
\newcommand*{\contentsim}[1]{ 
    \if #1S {
        \mathit{sim}_{\mathcal{#1}}
    }\else{
        \if #1D{
            \mathit{sim}_{\mathcal{#1}}
        }
        \else{
            \if #1L{
                \mathit{sim}_{\mathcal{#1}}
            }\else{
                \if #1O{
                    \mathit{sim}_{\mathcal{#1}}
                }\else{
                    \mathit{sim}_{#1}
                }\fi
            }\fi
        }\fi
    }\fi
}
\newcommand*{\weight}[1]{ 
    \if #1S {
        \alpha_{\mathcal{#1}}
    }\else{
        \if #1D{
            \alpha_{\mathcal{#1}}
        }
        \else{
            \if #1L{
                \alpha_{\mathcal{#1}}
            }\else{
                \if #1O{
                    \alpha_{\mathcal{#1}}
                }\else{
                    \alpha_{#1}
                }\fi
            }\fi
        }\fi
    }\fi
}
\newcommand*{\weightB}[1]{ 
    \if #1S {
        \beta_{\mathcal{#1}}
    }\else{
        \if #1D{
            \beta_{\mathcal{#1}}
        }
        \else{
            \if #1L{
                \beta_{\mathcal{#1}}
            }\else{
                \if #1O{
                    \beta_{\mathcal{#1}}
                }\else{
                    \beta_{#1}
                }\fi
            }\fi
        }\fi
    }\fi
}
\newcommand*{\nodelabel}[1]{
    \if #1N {
        \ell_{#1}
    }\else{
        \if #1Q {
            \ell_{#1}
        }\else{
            \if #1x{
                \ell_{#1}
            }\else{
                \ell_{\mathcal{#1}}
            }\fi
        }\fi
    }\fi
}
\newcommand*{\libraryset}[1]{\mathcal{L}_{#1}}

\begin{document}

\title{Similarity Search on Computational Notebooks}
\author{Misato Horiuchi\thanks{Graduate school of Information Science and Technology, Osaka University. \{horiuchi.misato, sasaki, chuanx, onizuka\}@ist.osaka-u.ac.jp} \\
\and
Yuya Sasaki\footnotemark[1]
\and
Chuan Xiao\footnotemark[1]
\and
Makoto Onizuka\footnotemark[1]}

\date{}

\maketitle

\begin{abstract}
Computational notebook software such as Jupyter Notebook is popular for data science tasks.
Numerous computational notebooks are available on the Web and reusable; however, searching for computational notebooks manually is a tedious task, and so far, there are no tools to search for computational notebooks effectively and efficiently.
In this paper, we propose a similarity search on computational notebooks and develop a new framework for the similarity search.
Given contents (i.e., source codes, tabular data, libraries, and outputs formats) in computational notebooks as a query, the similarity search problem aims to find top-$k$ computational notebooks with the most similar contents. 
We define two similarity measures; set-based and graph-based similarities.
Set-based similarity handles each content independently, while graph-based similarity captures the relationships between contents.
Our framework can effectively prune the candidates of computational notebooks that should not be in the top-$k$ results.
Furthermore, we develop optimization techniques such as caching and indexing to accelerate the search. 
Experiments using Kaggle notebooks show that our method, in particular graph-based similarity, can achieve high accuracy and high efficiency.  
\end{abstract}


\section{INTRODUCTION}
\label{sec:introduction}

Many data scientists currently use computational notebook software such as Jupyter Notebook and 
R Notebook
for data analytics.
They interactively conduct various data processing, for example, data cleaning, data mining, machine learning, and visualization.
Due to the increasing popularity of computational notebooks, numerous computational notebooks are available and reusable on the Web, such as GitHub and Kaggle~\cite{yan2020auto}.

We often search for computational notebooks to reuse them for our own data analytic tasks and to learn programming skills. 
When searching the computational notebook database, we would like to find computational notebooks to include source codes, tabular data, libraries, and/or output formats \emph{similar} to what we specify, because it is difficult to assign the search conditions (e.g., it may be equivalent to write codes) and there are no computational notebooks that perfectly match our conditions.
For example, we look for computational notebooks with analysis of data that is similar to ours, libraries, and functions that we would like to learn.

There is a great demand for a similarity search on computational notebooks.
Although some works partially support computational notebook search, there are no existing problem definitions and solutions to effectively and efficiently search for similar computational notebooks so far.
Existing methods for similar source code search~\cite{koschke2006clone, krinke2001identifying} and tabular data search~\cite{zhang2020finding} can be used for computational notebook search since their search targets are either source codes or tabular data, which are computational notebooks contents.
However, using them directly cannot meet our demand for computational notebook search. 
They only compute the similarity for their targets instead of computational notebooks as they are optimized for their search targets (i.e., source codes or tabular data). 
What they search for is only a subset of the contents in computational notebooks. 
Moreover, although combining their results to compute the similarity for computational notebooks is possible, it is inefficient because we need to conduct two searches individually.

Therefore, in this paper, we first propose a similarity search problem on computational notebooks and develop a new framework for the similarity search.
Our similarity search problem aims to find the top-$k$ computational notebooks with the most similar contents (i.e., source codes, tabular data, libraries, and output formats) to the contents specified by a given query. 
To measure a similarity between the given query and computational notebooks, we define two similarity measures; set-based and graph-based similarity.
Set-based similarity is a natural definition that computes the similarity of each content independently, while graph-based similarity aims to capture the relationships between contents.
In our graph-based similarity, that computational notebooks and queries are represented by directed acyclic graphs (DAGs) and computes the similarity based on subgraph matching.
The graph-based similarity search can finely specify user requirements.

We develop a new framework that can efficiently find the top-$k$ results based on the similarity definitions.
Our framework mainly targets graph-based similarity computation while we can use it for set-based similarity computation.
Our framework first converts computational notebooks into DAGs so that keep the execution flow and relationships between contents (e.g., codes read tabular data). 
In the search process, our framework employs subgraph matching 
to meet user requirements, and 
to effectively prune computational notebooks that should not be in the top-$k$ results.
The subgraph matching of DAGs requires less computational costs than computing the similarity scores, so it accelerates the similarity search.

To further accelerate the similarity search, our framework uses four optimization techniques; pruning, computation order optimizations, caching, and indexing.
These techniques drastically reduce the computation costs of both set-based and graph-based similarity computations.

In our experimental study, we evaluate that 
accuracy and efficiency
of our method by using $111$ computational notebooks shared on Kaggle. 
We show that our method achieves high $\mathit{nDCG}$~\cite{burges2005learning} through user evaluation that users manually measure the similarity of all the computational notebooks.
We show that our method is efficient and scalable by varying several settings such as $k$ and data size.
In particular, graph-based similarity is up to $157.7$ times faster than a baseline with a higher accuracy.

We summarize our contributions as follows:
\begin{itemize}
    \vspace{-1mm}
    \item We first propose the similarity search problems on computational notebooks. We define two similarity measures; set-based and graph-based similarity. The graph-based similarity search can effectively capture the user requirements (Section~\ref{sec:3_preliminary}).
    \vspace{-2.5mm}
    \item We develop a new framework that efficiently searches for similar computational notebooks (Section~\ref{sec:5_framework}).
    \vspace{-2.5mm}
    \item We accelerate the similarity search by using four optimization techniques; pruning, computation ordering, caching, and indexing (Section~\ref{sec:6_proposal}).%
    \vspace{-2.5mm}
    \item Through experimental evaluation using Kaggle notebooks, we show that our method is accurate, efficient, and scalable (Section~\ref{sec:7_experiment}).
\end{itemize}

The rest of this paper is organized as follows. Section~\ref{sec:2_relatedwork} describes related work. Section~\ref{sec:3_preliminary} explains our problem definition.
Section~\ref{sec:5_framework} presents the proposed search framework. 
Section~\ref{sec:6_proposal} presents optimization techniques of our search method.
The experimental results are shown in Section~\ref{sec:7_experiment}, and the paper is concluded in Section~\ref{sec:8_conclusion}.

\section{RELATED WORK}
\label{sec:2_relatedwork}
Computational notebooks are frequently used and published, yet there are some issues, for example, low reusability and difficulty to archive~\cite{chattopadhyay2020s}.
The similarity search on computational notebooks partially supports these issues. 
First, low reusability derives from insufficient description and meaningless file names.
This indicates that the keyword search is impractical due to their file name and descriptions. 
Second, archiving is hard because traditional source control systems are difficult to identify the actual changes between the notebooks computational notebooks. 
This is because computational notebooks are written by markup languages, and the systems detect all of these markup tag changes.
These issues can be partially solved by the similarity search on computational notebooks.
It helps to reuse them by retrieving computational notebooks with contents that users want instead of file names, and archive by identifying the old version notebooks that should have similar contents to new versions.
The similarity search is effective for more widely spreading usages of computational notebooks.

Many data scientists use tabular data in computational notebooks such as Pandas DataFrames, and researchers actively develop technologies focusing on tabular data in computational notebooks.
Zhang et al.~\cite{zhang2019juneau, zhang2020finding} proposed Juneau, a fast retrieval technique for tabular data in computational notebooks.
Besides, Yan et al.~\cite{yan2020auto} developed a system that automatically suggests the best operations to the tabular data by learning historical operations by data scientists, such as processing and transformation.
Chen et al. ~\cite{chen2019optimizing} proposed a method that reduces the time gap between the execution of Jupyter Notebooks' cells and the outputs of results. This method optimizes the execution of Pandas DataFrame to improve the interactivity of the Jupyter Notebook.
Among them, Juneau can be used for similarity search on computational notebooks and the others are not for searching. 
However, Juneau is not adapted to search for other contents of computational notebooks, such as source code and outputs since it can evaluate only the similarity of tabular data. Therefore, the technique for searching tabular data is not appropriate for computational notebooks themselves. We show that Juneau does not match user requirements in our experimental studies (see Figure~\ref{fig:ndcg} in Section~\ref{sec:7_experiment}).

Code similarity has been actively studied in the field of software engineering. A similar source code detection technique~\cite{krinke2001identifying} and a code clone detection technique~\cite{koschke2006clone} convert source code into directed graphs and searching for similar subgraphs. These techniques are not suitable for similarity search of computational notebooks because they specialize in source code and cannot target complex similarity computation (e.g., data similarity). In addition, they cannot handle the order of execution of cells and the relationship between contents.

\section{Similarity Search on Computational Notebooks}
\label{sec:3_preliminary}
    %
    
    There are no studies of similarity search on computational notebooks.
    After explaining computational notebooks, we propose similarity search problems on computational notebooks. 
    We then define two similarity measures; set-based and graph-based similarity.
    The difference between them is that the graph-based similarity aims to capture the relationships between contents instead that the set-based similarity handles each content independently.
    
    
    \subsection{Computational Notebook}
    A computational notebook is an editable document that consists of {\it cells} with source codes.
    The source codes in each cell are executed to import libraries, read tabular data to DataFrame, process/analyze the data, and output analytic results.
    The cells are executed one by one and never simultaneously.
    We define computational notebooks as follows:
    %

\begin{defi}[Computational notebook.]
    A computational notebook $N$ is a $5$-pair $\left(\mathcal{S}_N, \mathcal{D}_N, \mathcal{O}_N, \libraryset{N}, \mathcal{C}_N\right)$, where $\mathcal{S}_N$ is the source codes, $\mathcal{S}_N\neq \phi$, $\mathcal{D}_N$ is the set of tabular data, $\mathcal{O}_N$ is the multi-set of output formats, and $\libraryset{N}$ is the set of imported library names. 
    A cell $c_N$ is a block of a subset of $\mathcal{S}_N$. We denote the set of $c_N$ as $\mathcal{C}_N$.
\end{defi} 

We denote the set of $N$ as $\mathbb{N}$. 

\begin{figure*}[t]
    \begin{minipage}{0.225\hsize}
    \centering
    \includegraphics[width=1.0\linewidth]{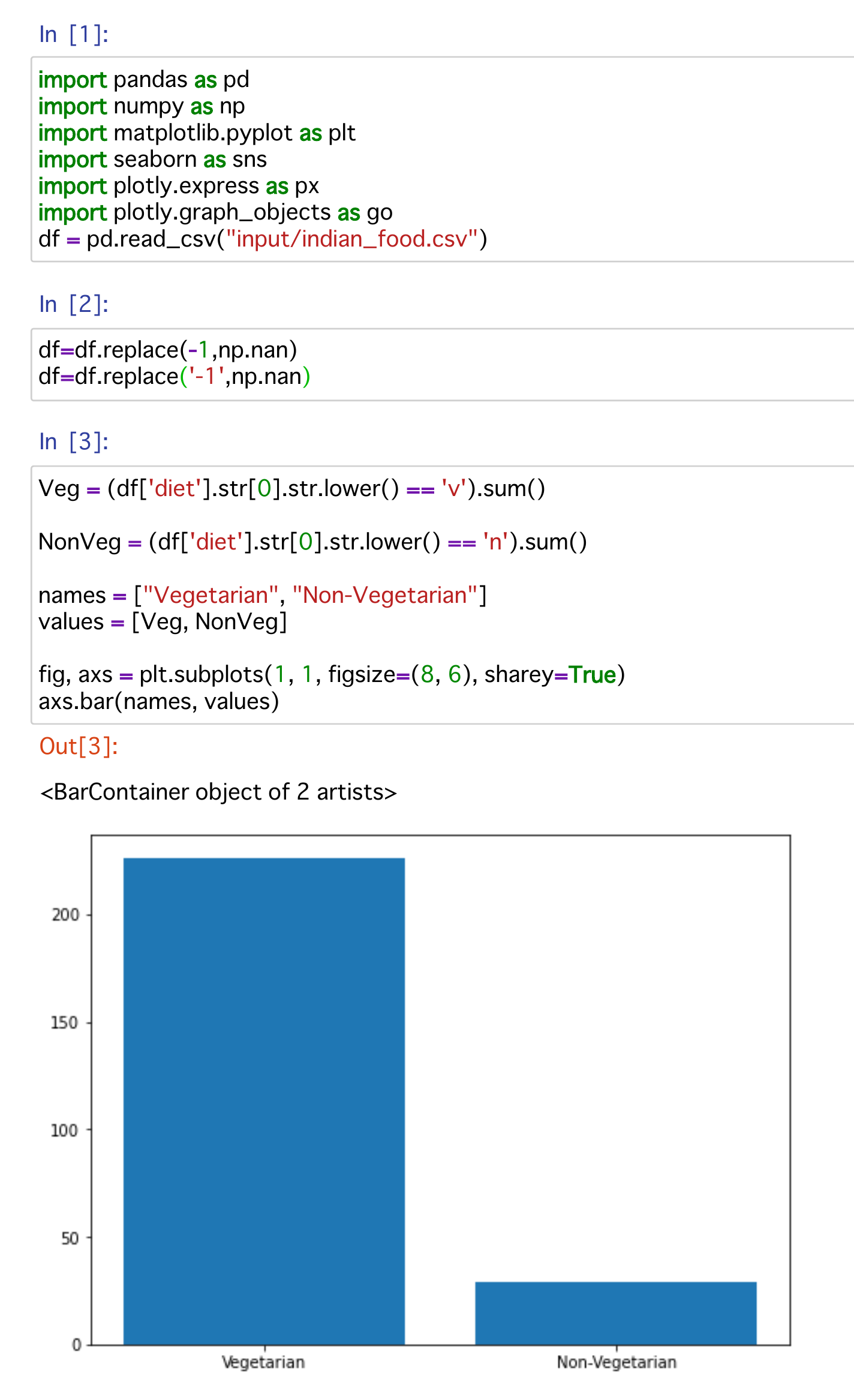}
    \caption{An example of computational notebooks}
    \label{fig:sample_notebook}
    \end{minipage}
    \begin{minipage}{0.75\hsize}
    \centering
    \includegraphics[width= 1.0\linewidth]{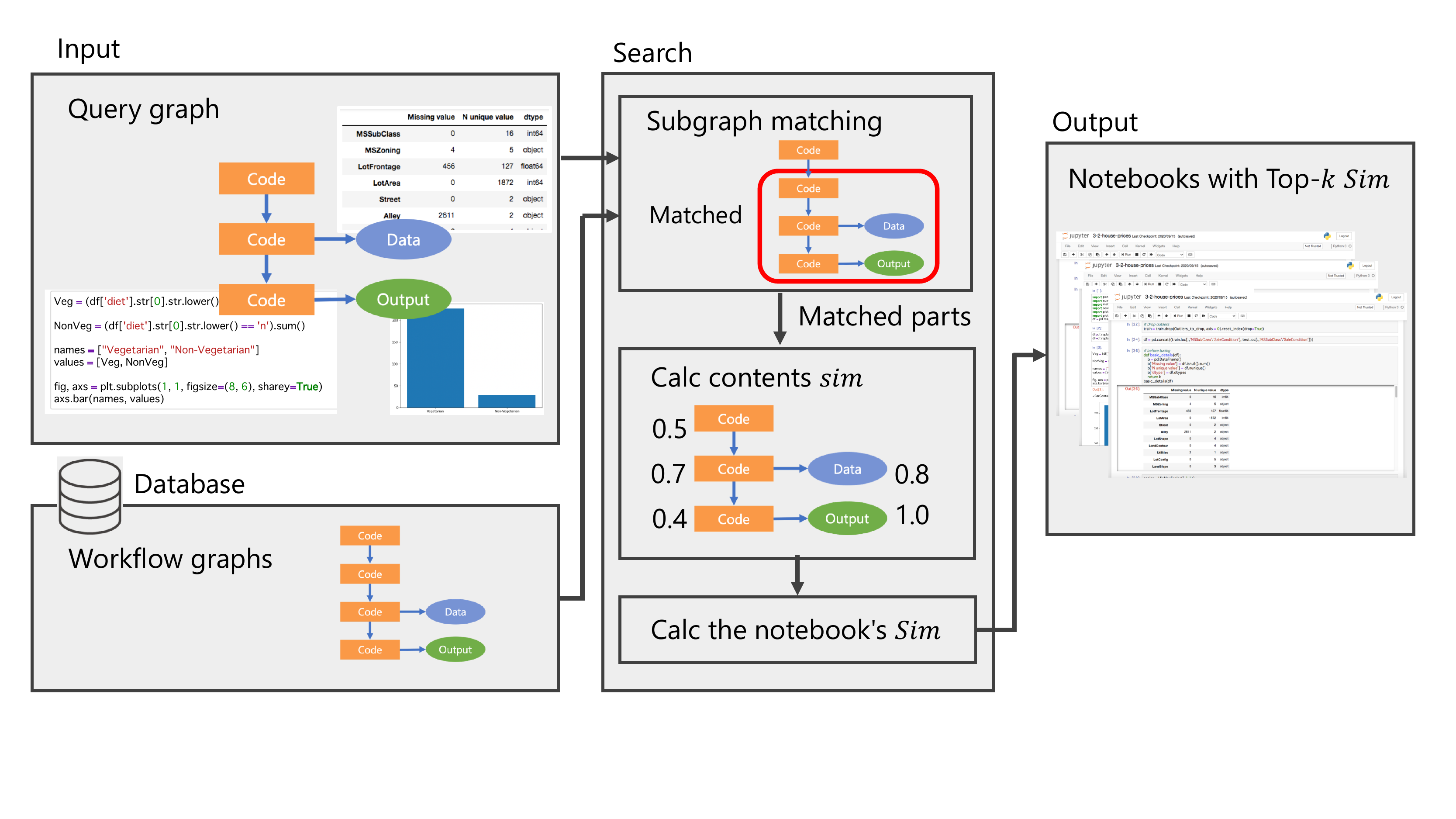}
    \vspace{-10mm}
    \caption{A framework of similarity search}
    \label{fig:framework}
    \end{minipage}
\end{figure*}

Figure~\ref{fig:sample_notebook} illustrates a computational notebook on Jupyter Notebook. This computational notebook has three cells; the first cell imports libraries and reads a CSV file, the second one conducts preprocessing for the tabular data, and the third one analyzes the data and outputs the analytic results.

\subsection{Problem Definition}

We define similarity of computational notebooks as the weighted sum of similarity of contents (e.g., source codes and tabular data). We first define {\it content similarity} as follows.

\begin{defi}[Content Similarity.] 
We define similarity measures of source codes, tabular data, output formats, and libraries as $\contentsim{S}\left(\mathcal{S}, \mathcal{S'}\right)$, $\contentsim{D}\left(\mathcal{D}, \mathcal{D'}\right)$, $\contentsim{O}\left(\mathcal{O}, \mathcal{O'}\right)$, $\contentsim{L}\left(\mathcal{L}, \mathcal{L'}\right)$, respectively.
The range of the similarity measure is $[0,1]$, and similar contents have the content similarity closer to one.
\end{defi}
We can use any similarity measures such as Jaccard similarity coefficient. 

We define our problem that we solve in this paper as follows:

\begin{problem}{\sc Similarity Search on Computational Notebooks.}
    Given a query $Q$, a set of computational notebooks $\mathbb{N}$, a natural number $k$, and non-negative weights that balance the importance of contents, we find a ranked list $\mathbb{A}$ of $k$ computational notebooks such that for $A \in \mathbb{A}$ and $N \in \mathbb{N} \backslash \mathbb{A}$, $\mathit{Sim}(Q,A) \geq \mathit{Sim}(Q,N)$.
\end{problem}

The ranked list depends on how to compute $\mathit{Sim}(Q,N)$.
We formulate two types of similarity: set-based and graph-based similarity.
Set-based similarity handles each content independently and computes the similarity between queries and computational notebooks by summing similarities of their entire contents.
On the other hand, graph-based similarity captures the relationships between the contents and then computes the similarity of contents per cells according to subgraph matching.

\subsubsection{Set-based Similarity}

The set-based similarity specifies contents as queries regardless of cells.
We define a query in set-based similarity as follows:

\begin{defi}[Query.]
A query $Q$ is a $4$-pair $\left(\mathcal{S}_Q, \mathcal{D}_Q, \mathcal{O}_Q, \libraryset{Q}\right)$, where each of them is source codes, a set of tabular data, output formats, and libraries, respectively.
\end{defi}

We define the set-based computational notebook similarity as follows:

\begin{defi}{\sc Set-based Computational Notebook similarity.}
\label{def:set_based_notebook_similairity}
    Given a query $Q$, a computational notebook $N$, we define a similarity measure between $Q$ and $N$ as $\mathit{Sim}(Q,N)$:
    \begin{align}
        \mathit{Sim}(Q,\!N) \!\!=\!\! \weight{S} \contentsim{S}\!\left(\mathcal{S}_Q,\! \mathcal{S}_N\right)\!\!
        +\!\! \weight{D} \contentsim{D}\! \left(\mathcal{D}_Q,\! \mathcal{D}_N\right)  \\
        \;\;\;\;\;\; + \weight{O} \contentsim{O}\!\left(\mathcal{O}_Q,\! \mathcal{O}_N\right)\!\! +\!\! \weight{L} \contentsim{L}\! \left(\libraryset{Q},\! \libraryset{N}\right)  \nonumber
    \end{align}
    where $\weight{S}$, $\weight{D}$, $\weight{O}$, $\weight{L}$ are weights corresponding to importance of source codes, tabular data, output formats, and libraries, respectively.
\end{defi}

This similarity search problem can be solved by computing the sum of content similarity.




\subsubsection{Graph-based Similarity}
\label{sec:4_graph_based_search}


The set-based similarity does not capture the execution orders of cells and relationships between contents.
We model computational notebooks and queries as directed acyclic graphs (DAGs) to capture them.
We define {\it workflow graphs} for computational notebooks and a {\it query graph}, respectively.


\begin{defi}[Workflow graph.]
    The execution flow and contents relationships of a computational notebook $N$ is represented by a {\rm DAG} $\workflowgraph=\left(V_N,E_N,\nodelabel{N},\workflowattr,\libraryset{N}\right)$, and we call $\workflowgraph$ workflow graph. $V_N$ is a set of nodes, $E_N \subseteq V_N \times V_N$ is a set of directed edges $e(v, v')$ going from $v\in V_N$ to $v'\in V_N$, $\nodelabel{N}$ is a mapping from each node to a label, 
    $\workflowattr$ is a mapping from each node to a set of attributes.
    For $v\in V_N$, $\nodelabel{N}(v)$ and $\workflowattr(v)$ are the label and the attribute of $v$, respectively. As node labels in the workflow graph, the node corresponding to the source codes of a cell is labeled $\nodelabel{S}$, the node corresponding to tabular data is labeled $\nodelabel{D}$, and the node corresponding to the format of an output of a cell is labeled $\nodelabel{O}$. Therefore, $\nodelabel{N}:V_N \rightarrow \{ \nodelabel{S}, \nodelabel{D}, \nodelabel{O}\}$. The node $v$ has the attribute $\workflowattr(v)$ corresponding to the label $\nodelabel{N}(v)$. If $\nodelabel{N}(v)$ is $\nodelabel{S}$, $\nodelabel{D}$, and $\nodelabel{O}$, then $\workflowattr(v)$ is a set of source codes, tabular data, and output formats, respectively.
\end{defi}
Note that one computational notebook corresponds to one workflow graph.
We define $\mathbb{W}$ as the set of workflow graphs.

A query is also represented by a DAG that consists of a set of the source codes, tabular data, libraries, and output formats.
We define a query graph as follows:



\begin{defi}[Query Graph.]
   A query graph $Q$ is a $5$-pair $\left(V_Q,E_Q,\nodelabel{Q},\queryattr,\libraryset{Q}\right)$ as well as a workflow graph. The node labels of the query graph are $\{ \nodelabel{S}, \nodelabel{D}, \nodelabel{O}, *\} $. Label $*$ denotes the reachability label, indicating that there is a path between its adjacent nodes. 
   We note that nodes with label $*$ are not adjacent.
\end{defi}

The nodes in workflow and query graphs represent source codes in cells, tabular data, and outputs.
Thus, we can specify our requirements with finer granularity than the set-based similarity.
The difference between workflow and query graphs is that query graphs can have reachability as labels, which enables the flexibility to specify the relationship between contents.


To compare the similarity between workflow and query graphs, we consider correspondences between nodes in workflow and query graphs.
For this purpose, we define subgraph matching from query graphs to workflow graphs as follows:

\begin{defi}[Subgraph Matching.]
Given a query graph $Q=\left(V_Q,E_Q,\nodelabel{Q},\queryattr,\libraryset{Q}\right)$ and a workflow graph $\workflowgraph=\left(V_N,E_N,\nodelabel{N},\workflowattr,\libraryset{N}\right)$, we define a {\it mapping} $M$ from $Q$ to $\workflowgraph$ as follows: 
    \vspace{-1mm}
    \begin{enumerate}
        \item If $\nodelabel{Q}(v)\neq *$, then $v$ corresponds to a node $u\in V_N$.\vspace{-3mm}
        \item For $\forall v \in V_Q$, if $\nodelabel{Q}(v)\neq *$, then $\nodelabel{Q}(v)=\nodelabel{N}(M(v))$.\vspace{-3mm}
        \item If an edge $e(v_i, v_j)\in E_Q$ exists, $\nodelabel{Q}(v_i)\neq *$, and $\nodelabel{Q}(v_j)\neq * $, then $\exists e; e(M(v_i), M(v_j))\in E_N$.\vspace{-3mm}
        \item If edges $e(v_i, v_m)$ and $e(v_m, v_j)$ exists,  $\nodelabel{Q}(v_i)\neq *$, $\nodelabel{Q}(v_j)\neq *$, and $\nodelabel{Q}(v_m)=*$, then there is a path from $M(v_i)$ to $M(v_j)$.\vspace{-3mm}
    \end{enumerate}
\end{defi}
There are multiple mapping from $Q$ to $\workflowgraph$. 
We denote a set of mapping as $\mathbb{M}$.

Next, we define the similarity measure between $Q$ and $\workflowgraph$ as follows:

\begin{defi}{\sc Graph-based Similarity.}
\label{def:graph_based_notebook_similarity}
    Given a query graph $Q$, a workflow graph $\workflowgraph$, and a set of mappings $\mathbb{M}$, we define similarity between $Q$ and $\workflowgraph$ for $M\in \mathbb{M}$ as $\mathit{Sim}(Q,\workflowgraph,M)$:
    \begin{align}
        & \mathit{Sim}(Q,\workflowgraph,M) = \weightB{L} \contentsim{L} \left(\libraryset{Q}, \libraryset{N}\right) \\
        & \;\;\;\;\; + \sum_{v\in V_Q} \weightB{v} \contentsim{v} \left( \queryattr(v),\workflowattr(M(v))\right)
         \nonumber
    \end{align}
    where  $\weightB{v}$ are $\weightB{S}$, $\weightB{D}$, $\weightB{O}$ and $\contentsim{v}$ are $\contentsim{S}$, $\contentsim{D}$, $\contentsim{O}$, for $\nodelabel{Q}(v)=\nodelabel{S}, \nodelabel{D}, \nodelabel{O}$, respectively.
    For $\x \in \{\mathcal{S}, \mathcal{D},\mathcal{O}\}$, we denote $\weightB{\x}=\frac{\weight{\x}}{\mathit{size}_{\x}(V_Q)}$ if $\mathit{size}_{\x}(V_Q) \neq 0$ and $\weightB{\x}=0$ if $\mathit{size}_{\x}(V_Q) = 0$,
    where $\mathit{size}_{\x}(V)$ is the number of nodes with label $\nodelabel{x}$ in a set of nodes $V$, and $\weightB{L}=\weight{L}$.
    We define a similarity between $N$ and $Q$ as $\mathit{Sim}(Q,N)$. If there are multiple mappings, the highest $\mathit{Sim}(Q,\workflowgraph,M)$ is $\mathit{Sim}(Q,N)$ as follows:
    \begin{align}
        \label{eq:計算ノートブックの類似度}
        \!\!\! \mathit{Sim}(Q,\!N)\! =\! 
        \begin{cases}
             \max_{M\in \mathbb{M}} \mathit{Sim}\left(Q,\!\workflowgraph,\!M \right)\! &\!\! \mathbb{M}\! \neq\! \phi \\
            0\! &\!\! \mathbb{M}\! = \!\phi.
        \end{cases}
    \end{align}
\end{defi}


We note that $\weightB{S}$, $\weightB{D}$, $\weightB{L}$, and $\weightB{O}$ are for the normalization of the weights since 
the maximum value of sum of each content similarity
increases if the number of nodes increases.
\section{Our Framework}
\label{sec:5_framework}

    In this section, we describe the proposal similarity search framework. We mainly focus on graph-based similarity while our framework can be used for set-based similarity.
    
    Figure~\ref{fig:framework} illustrates our framework architecture.
    We input a query and weights of each content similarity. 
    The workflow graphs of computational notebooks are stored in the database.
    Our framework computes similarity between the query and workflow graphs, and outputs the top-$k$ computational notebooks with the highest similarity.
    The search part of the framework consists of two major components: subgraph matching and similarity computation.
    
    We first describe how to construct the workflow graphs, and then we explain a search method to compute the top-$k$ computational notebooks.

    \subsection{Workflow Graph Construction}
    \label{sec:Building_workflow_graphs}
    
        Our graph-based similarity computes the similarity between a given query and computational notebooks to compare the execution orders of cells and the relationships between contents.
        For this purpose, we transform computational notebooks into graphs with maintaining the workflow of cells.
        We describe how to transform a computational notebook $N$ into a workflow graph $\workflowgraph$ in the following.
                
        A node set $V_N$ represent cells in $\mathcal{C}_N$, tabular data in $\mathcal{D}_N$, and outputs in $\mathcal{O}_N$. Each of their labels is $\nodelabel{S}$, $\nodelabel{D}$, and $\nodelabel{O}$, respectively. Node attributes $\workflowattr(v)$ are any of $\mathcal{S}_N$ 
        , $\mathcal{D}_N$, and $\mathcal{O}_N$ corresponding to the node labels.
    
        Constructing an edge set $E_N$ is as follows. $E_N$ includes edges between nodes with $\nodelabel{S}$ and between nodes with $\nodelabel{S}$ and $\nodelabel{D}$ or $\nodelabel{O}$.
        If $v_i, v_j \in V_N$ corresponding to the source codes $\mathcal{S}_N$ and source codes of $v_i$ are immediately executed after that of $v_j$, there is an edge $v_i$ to $v_j$.
        A node $v_d \in V_N$ with label $\nodelabel{D}$ has an edge $e(v_i, v_d)$ if its tabular data is stored into a variable in the source codes of the cell corresponding to $v_i$.
        Furthermore, if the source codes of the cell that uses the variable corresponding to $v_d$ exists, there is an edge $e(v_d, v_j)$.
        For the node $v_o\in V_N$ with label $\nodelabel{O}$, we add edge $e(v_i, v_o)$ if the source codes corresponding to $v_i$ outputs.
        
        In these processes, we can maintain the execution flow and contents relationships of computational notebooks  without loss of any information. 

    \subsection{Search Method}
    \label{sec:Naive_method}
        Our search method computes similarities between queries and computational notebooks one by one.
        The search method is as follows:

        \begin{enumerate}
            \vspace{-2mm}
            \item It enumerates all matches, which are the results of subgraph matching between $Q$ and all workflow graphs $\mathbb{W}$.
            \vspace{-2.5mm}
            \item 
            It calculates the content similarity $\contentsim{L}(\libraryset{Q}, \libraryset{N})$ and $\contentsim{v} \left( \queryattr(v),\workflowattr(M(v))\right)$, $\forall v\in V_Q$ for   $M\in \mathbb{M}$. Then, after calculating $\mathit{Sim}(Q,\workflowgraph,M)$, it calculates $\mathit{Sim}(Q,N)$.\vspace{-2.5mm}
            \item It repeats the above processes until obtaining the similarities of all computational notebooks. Then, it returns the $k$ computational notebooks with the highest $\mathit{Sim}(Q,N)$.
        \end{enumerate}
        
        This search method is simple and has many opportunity to be accelerated.
        In the next section, we develop optimization techniques for accelerating this search method.
\section{Search Optimization}
\label{sec:6_proposal}

    Our search method has many opportunity to improve the efficiency. 
    First, it calculates all the content similarity, even when computational notebooks are obviously not in the top-$k$ results. 
    Second, it computes the content similarity of all the computational notebooks regardless of their computational cost. 
    Third, it computes the similarity of the same contents multiple times.
    Finally, it performs subgraph matching even for workflow graphs that do not match obviously.

    In this section, we present four optimization techniques to support the above opportunities.
    Our optimization techniques are (1) pruning candidates of computational notebooks, (2) optimization calculation orders of content similarity, (3) caching content similarity, and (4) topology-based indexing.

    \subsection{Pruning computational notebooks and content similarity computation}
    \label{sec:計算済類似度との比較による枝刈り}
        We prune computational notebooks and content similarity computations by comparing the similarity that is computed already. The general idea is that we prune computational notebooks if their possible maximum similarity are not larger than the current $k$-th similarity. 
        
        Suppose that some but not all of the content similarity $\contentsim{L}\left(\libraryset{Q}, \libraryset{N}\right)$ and $\contentsim{v} \left( \queryattr(v),\workflowattr(M(v))\right)$ have already been computed. 
        Let $\mathit{MaxSim}\left(Q, \workflowgraph, M\right)$ be the possible maximum value as follows:
        \begin{align}
            &\mathit{MaxSim}\left( Q, \workflowgraph, M\right) = \weightB{L} \contentsim{L} \left(\libraryset{Q}, \libraryset{N}\right) \\
            &\;\;\;\; + \sum_{v\in V'_Q } \weightB{v} \contentsim{v} \left( \queryattr(v),\workflowattr(M(v))\right)  + \!\!\!\!\!\sum_{u\in V_Q\setminus V'_Q }\!\!\!\!\! \weightB{u}. \nonumber
        \end{align}
        where  $V'_Q \subset V_Q$ is the set of nodes that their content similarity are already computed. We note that maximum content similarity is one.

        Given the $k$-th similarity $\mathit{Sim}_k$, we can prune the computational notebooks if it holds the following:
        \begin{align}
            \mathit{Sim}_k >  \mathit{MaxSim}\left( Q, \workflowgraph, M\right).
        \end{align}

        Furthermore, we prune the content similarity computation by comparing them with the computed similarity of the same workflow graph. 
        Let the mapping set which is computed the similarity be $\mathbb{M'}\subseteq \mathbb{M}$ and
        \begin{align}
            \max_{M'\in \mathbb{M'} }\mathit{Sim}\left( Q,\workflowgraph,M'\right) >  \mathit{MaxSim}\left( Q, \workflowgraph, M\right)
        \end{align}
        holds, then Equation (\ref{eq:計算ノートブックの類似度}) clearly shows that this similarity $\mathit{Sim}\left( Q, \workflowgraph, M\right)$ cannot be the  $\mathit{Sim}(Q,N)$. 

        These pruning allow us to reduce similarity computations that are clearly not in the top-$k$ or the largest similarity value in the workflow graph without computing the similarity for all the constituent nodes.

    \subsection{Ordering contents similarity computation}
    \label{sec:関連度計算順序の最適化}
    
        We can improve the effectivity of pruning by optimizing the order of content similarity computation.
        Given the computation costs of content similarity, we prioritize the contents similarity computation with small costs to reduce the content similarity computation with the largest cost as much as possible. 
        
        In this optimization, we first all content similarity except for the one with the largest cost (we use $\nodelabel{\theta}$ as the label with largest cost) and then obtain tentative similarities of computational notebooks. We sort a pair of $(W, M)$ in descending order of the tentative similarity.
        After these processes, we compute the exact similarity from the highest tentative similarity with pruning the computational notebooks and content similarity computations in the way of Section~\ref{sec:計算済類似度との比較による枝刈り}.
          
        We note that the computation cost of subgraph matching is smaller enough than that of content similarity computation from our preliminary experiments. 

    \subsection{Caching computed content similarity}  
    \label{sec:計算済み関連度のキャッシング}

We can reduce the wastefulness of computing the similarity between the same nodes multiple times by caching computed similarity.
We store the contents similarity once they are calculated. We reuse the stored contents similarity to reduce computation costs.

  \subsection{Topology-based indexing}
  \label{sec:インデックスの構築}
      We prune the computational notebooks by using the topological information of the workflow graph.
      The maximum incoming degree of an edge in $W$ is $d_{\mathit{in}}(W)$, and the maximum outgoing degree is $d_{\mathit{out}}(W)$.
      Before subgraph matching of $Q$ to $\workflowgraph$, we know that there are no matches if either of the following holds.
      \begin{itemize}
        \vspace{-1mm}
        \item $\exists \x\in \{\mathcal{S}, \mathcal{D}, \mathcal{O}\}$, $\mathit{size}_{\x}(V_N) < \mathit{size}_{\x}(V_Q)$
        \vspace{-2.5mm}
        \item $d_{\mathit{in}}(\workflowgraph) < d_{\mathit{in}}(Q)$
        \vspace{-2.5mm}
        \item $d_{\mathit{out}}(\workflowgraph) < d_{\mathit{out}}(Q)$
      \end{itemize}
      We can prune workflow graphs that do not obviously match to queries.

\subsection{Algorithm}
\label{sef:algorithm}

\begin{algorithm}[!t]
    \caption{Graph-based similarity search with optimization techniques}
    \label{algorithm}
	{\small
        \SetKwInOut{Input}{input}
        \SetKwInOut{Output}{output}
        \Input{Query graph $Q$, set of workflow graphs $\mathbb{W}$, natural number $k$, non-negative numbers $\weightB{S}, \weightB{D}, \weightB{O}, \weightB{L}$, index, a node label $\nodelabel{\theta}$}
        \Output{$k$ computational notebooks with the $k$ highest $\mathit{Sim}(Q, N)$}
        \For{$\workflowgraph \in \mathbb{W}$}{
            \If{$Q$ has no matches in $\workflowgraph$ based on the index}{
                $\mathbb{M}[\workflowgraph] \leftarrow \phi$\\
            }\Else{
                $\mathbb{M}[\workflowgraph] \leftarrow$ Subgraph Matching$(Q,\workflowgraph)$\\
            }
            $\mathit{Sim}(Q, N) \leftarrow 0$\\
            \For{$M \in \mathbb{M}[\workflowgraph]$}{
                $\mathit{Sim}(Q, \workflowgraph, M) \leftarrow \weightB{L}\contentsim{L}(\libraryset{Q}, \libraryset{N})$\\
                \For{$v \in V_Q$}{
                    \If{$\nodelabel{Q}(v)\neq \nodelabel{\theta}$}{
                        $\mathit{Sim}(Q, \workflowgraph, M) \leftarrow \mathit{Sim}(Q, \workflowgraph, M) + \weightB{v} \contentsim{v} \left( \queryattr(v),\workflowattr(M(v))\right)$\\
                    }
                }
            }
        }
        Sort $(\mathbb{W},\mathbb{M})$ by $\mathit{Sim}(Q, \workflowgraph, M)$\\
        \For{$(\workflowgraph,M) \in (\mathbb{W},\mathbb{M})$}{
                \For{$v \in V_Q$}{
                    \If{$\nodelabel{Q}(v)=\nodelabel{\theta}$}{
                        $\mathit{Sim}(Q, \workflowgraph, M) \leftarrow \mathit{Sim}(Q, \workflowgraph, M) + \weightB{\theta} \contentsim{\theta}(\queryattr(v),\workflowattr(M(v)))$\\
                        Update $\mathit{MaxSim}\left( Q, \workflowgraph, M\right)$\\
                        \If{$\mathit{MaxSim}\left( Q, \workflowgraph, M\right) < \mathit{Sim}_k$ or $\mathit{MaxSim}\left( Q, \workflowgraph, M\right) < \mathit{Sim}(Q, N)$}{
                            {\bf break}\\
                        }
                    }
                }
                \If{$\mathit{Sim}(Q, N) < \mathit{Sim}(Q, \workflowgraph, M)$}{
                    $\mathit{Sim}(Q, N) \leftarrow \mathit{Sim}(Q, \workflowgraph, M)$\\
                }
                $\mathbb{A} \leftarrow$ sort $\mathbb{N}$ by $\mathit{Sim}(Q, N)$\\
                $\mathit{Sim}_k \leftarrow \mathit{Sim}(Q, \mathbb{A}[k-1])$\\
        }
        \KwRet{$\mathbb{A}[0...k-1]$}\\
    }
\end{algorithm}

Algorithm~\ref{algorithm} shows the pseudo-code of the graph-based similarity search with optimization techniques.
In this algorithm, $\mathbb{M}[\workflowgraph]$ represents the set of mappings between the workflow graph $\workflowgraph$ and the query $Q$. A node label $\nodelabel{\theta}\in \{\nodelabel{S}, \nodelabel{D}, \nodelabel{O}\}$ is the label corresponding to the heaviest computation cost similarity.
All content similarity are computed with caching.
First, this algorithm computes subgraph matching using the indexes and similarity except $\nodelabel{\theta}$ similarity (lines $1$--$11$).
Then, it computes the $\nodelabel{\theta}$ similarity and the similairty of computational notebooks while pruning computational notebooks and content similarity (lines $12$--$23$).
Finally, it returns the top-$k$ computational notebooks (line $24$).



\section{EXPERIMENTAL EVALUATION}
\label{sec:7_experiment}
In this section, we show the experimental evaluation to show accuracy and efficiency of our methods.
We assume that computational notebook similarity search is generally run on client computers.
Thus, in our experiment, we use a Mac notebook with 2.30GHz Intel Core i5 and 16GB memory. 
We implement and execute all algorithms by Python.
We use PostgreSQL, Neo4J and SQLite to store tabular data,  workflow graphs, and saved query graphs, respectively.

\subsection{Experimental Setting}
\label{sec:実験環境}
We describe the dataset, comparison methods, and queries that we used in our experimental study.

\noindent
{\bf Dataset. }
We use $111$ computational notebooks shared on Kaggle as the dataset.
They are used for various purposes, including machine learning, data analysis, and data visualization.

We convert them into workflow graphs as execution order of cells is from top to bottom, since most of the computational notebooks on Kaggle are usually supposed to run cells in that order. 
In addition, we obtain the contents of the tabular data by executing all the cells in the order and saving into the database. 
The time for converting computational notebook into workflow graph was as long as the running time of computational notebooks for tabular data collection because the graph construction time is shorter enough to be ignored than the data collection time.
Tables~\ref{tab:sta_of_dataset}--\ref{tab:sta_of_dataset_3} show statistics of workflow graphs, source codes, and tabular data, respectively.

        
\noindent
{\bf Measures of content similarity. }
We describe the measures of content similarity in our experiments.

(1) Source code.
We separate the strings of source codes in the cells into a set of words.
The delimiters are space, new line, period, and equal.
We define the similarity between two sets of source codes as the Jaccard similarity coefficient for the sets of words.

(2) Tabular data.
Given two tabular data $D_A$ and $D_B$, we separate them into columns, we define each of $\mathit{col}_A$ and $\mathit{col}_B$ as a set of unique values per columns.
We define the number of columns in $D_A$ is $|D_A|$, and it is fewer than the number of rows in $D_B$. We define $\mathit{Jacc}(\mathit{col}_A, \mathit{col}_B)$ as the similarity between $\mathit{col}_A$ and $\mathit{col}_B$ which is calculated as Jaccard similarity coefficient. 
Then we select $|D_A|$ pairs in descending order to $\mathit{Jacc}(\mathit{col}_A, \mathit{col}_B)$.
Injection $T_{\mathit{col}}:\mathit{col}_A \rightarrow \mathit{col}_B$ explains correspondence of the pairs, we calculate the similarity of the table by $\frac{1}{s}\sum_{\mathit{col}_A \in D_A}{\mathit{Jacc}(\mathit{col}_A, T_{\mathit{col}}(\mathit{col}_A))}$.

(3) Library.
We use the Jaccard similarity coefficient to sets of words of library names used in the computational notebooks. 

(4) Output.
We use similarity of outputs as whether two outputs match in the type {\rm \{DataFrame, text, png\}}. The similarity is one if they match, and zero if they do not match. 

        
\begin{table}[t]
\centering
\footnotesize
\caption{Statistics of workflow graphs}
\vspace{-2.5mm}
\label{tab:sta_of_dataset}
    \begin{tabular}{lcccc} \hline
        Elements & Max & Min & Average & Total \\  \hline
        \# of nodes with $\nodelabel{S}$ & 123 & 5 &29.92 & 3321\\ 
        \# of nodes with $\nodelabel{D}$ & 44 & 0 & 8.86 & 983 \\ 
        \# of nodes with $\nodelabel{O}$ & 107 & 0 & 26.59 & 2952 \\ 
        \# of edges & 282 & 12 & 73.10 & 8114 \\ 
        Max in-degrees & 8 & 1 & 2.71 & - \\ 
        Max out-degrees & 21 & 2 & 4.97 & - \\ 
        \# of libraries & 22 & 2 & 6.86 & 762 \\
        \# of DataFrame & 43 & 1 & 4.28 & 475 \\
        \# of png & 46 & 1 & 8.79 & 976 \\
        \# of text & 47 & 1 & 13.52 & 1501 \\ \hline
    \end{tabular}

\vspace{5mm}
\centering
\footnotesize
\caption{Statistics for lines of source codes} 
\label{tab:sta_of_dataset_2}
\begin{tabular}{ccccc} \hline   
    Max & Min & Per cell &  Per notebook & Total \\ \hline
    9043 & 1 & 232.52 & 48261.63 & 772186 \\ \hline
\end{tabular}

\vspace{5mm}
\centering
\footnotesize
\caption{Statistics for tabular data} 
\label{tab:sta_of_dataset_3}
\begin{tabular}{cccc} \hline   
    Max [MB]  & Min [Byte] & Average [MB] & Total [MB]  \\ \hline
    248.7 & 144 & 85.9 & 9538.9 \\ \hline
\end{tabular}

\end{table}

        \begin{figure*}[!t]
            \centering
            \begin{minipage}{0.3\hsize}
                \centering
                \includegraphics[width=\hsize]{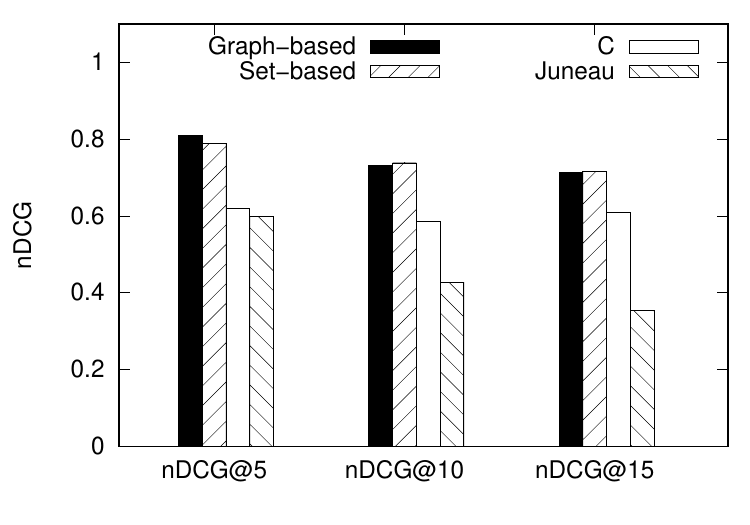}
                \caption{Search accuracy}
                \label{fig:ndcg}
            \end{minipage}
            \begin{minipage}{0.3\hsize}
                \centering
                \includegraphics[width=\hsize]{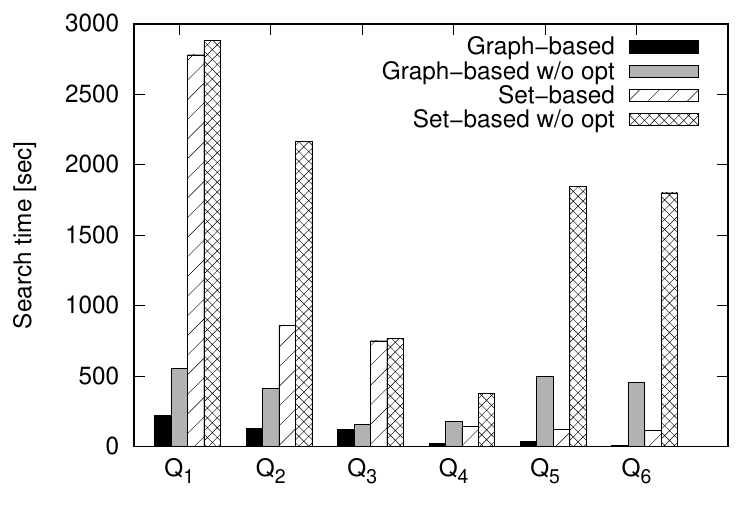}
                \caption{Top-$10$ search time}
                \label{fig:overview3_k10}
            \end{minipage}
            \begin{minipage}{0.3\hsize}
                \centering
                \includegraphics[width=\hsize]{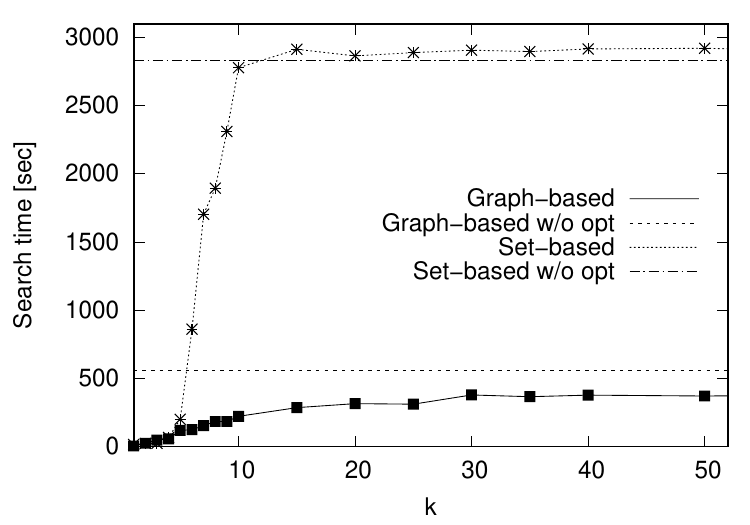}
                \caption{Search time on rank $k$}
                \label{fig:running_time_query2_3}
            \end{minipage}
        \end{figure*}
        
\noindent
{\bf Comparison methods. }
We use six methods for evaluation: {\sf Graph-based}, {\sf Graph-based w/o opt}, {\sf Set-based}, {\sf Set-based w/o opt}, 
{\sf C}, and {\sf Juneau}.
{\sf Graph-based} and {\sf Graph-based w/o opt} are the graph-based similarity methods with and without all optimization techniques, respectively. 
{\sf Set-based} is the set-based similarity method with optimization techniques mentioned mentioned in Sections~\ref{sec:計算済類似度との比較による枝刈り} and \ref{sec:関連度計算順序の最適化}. Other optimization techniques cannot apply to the set-based similarity.
{\sf Set-based w/o opt} does not use optimization techniques and can be considered as a baseline that independently uses similarity measures for each content.
{\sf C} and {\sf Juneau} use only code similarity and data similarity, respectively. 
{\sf Juneau} is a data search method~\cite{zhang2020finding}, which approximately finds similar tabular data.

\noindent
{\bf Queries and parameters.}
We use six queries whose query graphs have different attributes for each node.
$Q_1$--$Q_3$, and $Q_4$--$Q_6$ are the same topology, respectively.
Each of them is a fragment of a computational notebook in the dataset.

In set-based similarity, we set ($\weight{S}$, $\weight{D}$, $\weight{L}$, $\weight{O}$) as (32, 2, 1, 1) for $Q_1$--$Q_3$, and (32, 1, 1, 1) for $Q_4$--$Q_6$.
In graph-based similarity, we set ($\weightB{S}$, $\weightB{D}$, $\weightB{L}$, $\weightB{O}$) as (8, 1, 1, 1) for all queries.
In both similarity, $\nodelabel{\theta}$ is $\nodelabel{D}$ for all queries. 
    
\subsection{Results and Analysis}
\label{sec:実験結果}
We show the experimental results to validate the accuracy and efficiency of our method. 

    
\subsubsection{Accuracy}
\label{sec:検索精度}
We evaluate the accuracy of search results based on user experiments.
In the user experiments, nine users manually scored similarity of computational notebooks to $Q_4$--$Q_6$.
The score range is from $1$ to $5$, and we use the average of each score of the manual evaluations for the queries.



Figure~\ref{fig:ndcg} shows nDCG~\cite{burges2005learning}  of {\sf Graph-based}, {\sf Set-based}, {\sf C}, and {\sf Juneau}. 
{\sf Graph-based} and {\sf Set-based} achieve high nDCG.
Since {\sf Graph-based} is more accurate than {\sf Set-based},  we can confirm that it is effective to preserve the computation order of cells and relationships of contents.
From these results, we can see that our problem definitions are useful for the similarity search.
        
    \subsubsection{Efficiency}
    \label{sec:検索時間}

We evaluate the efficiency of our methods.

Figure~\ref{fig:overview3_k10} shows the search time for the Top-$10$ search.
The search time of {\sf Graph-based} is much smaller than that of {\sf Set-based}.
This indicates that subgraph matching is effective to reduce the computation costs without sacrificing the accuracy.
Optimization techniques can accelerate the search method for both  {\sf Graph-based} and {\sf Set-based}.

Figure~\ref{fig:running_time_query2_3} shows the search time with varying $k$.
For all $k$, {\sf Graph-based} is the most efficient among all the methods.
In particular, when $k$ is small, the effectiveness of optimization techniques is large.
This is because pruning works well due to high $k$-th similarity.
The search time of methods without optimization techniques is constant because they compute all computational notebooks without pruning.

\section{Conclusion}
\label{sec:8_conclusion}
In this paper, we defined similarity search on computational notebooks and  proposed a new framework for searching computational notebooks that comprehensively considers source codes, tabular data, libraries, and outputs.
We defined two similarity measures; set-based and graph-based similarity.
The set-based similarity handles contents independently, while graph-based similarity captures the relationships between contents.
Our framework based on graph-based similarity with optimization techniques efficiently finds the top-$k$ similar computational notebooks.
Through our experiments using Kaggle notebooks, we showed that the graph-based similarity achieves high accuracy and our methods are efficient.

As the future work, we aim to improve our similarity search to flexibly specify user requirements such as diverse results and dissimilar contents.
In addition, it is interesting to investigate the relationships between tabular data and source codes.

\bibliographystyle{abbrv}
\bibliography{bibliography}

\end{document}